\documentclass[12pt,preprint]{aastex}
\usepackage{emulateapj5}
\newif\ifAMStwofonts
\AMStwofontstrue

\def\kms{km~s$^{-1}$}

\def\ga{\mathrel{\hbox{\rlap{\hbox{\lower4pt\hbox{$\sim$}}}\hbox{$>$}}}}
\def\la{\mathrel{\hbox{\rlap{\hbox{\lower4pt\hbox{$\sim$}}}\hbox{$<$}}}}

\shorttitle{Does TSAS exist in the ISM ? }

\shortauthors{S.\ Stanimirovi\'{c} et al.}

\begin{document}

\title{Does tiny-scale atomic structure exist in the interstellar medium ?}

\author{S. Stanimirovi\'{c}}
\affil{Radio Astronomy Lab, UC Berkeley, 601 Campbell Hall,
Berkeley, CA 94720}
\email{sstanimi@astro.berkeley.edu}
\author{J. M. Weisberg, A. Hedden, K. E. Devine,J. T. Green}
\affil{Department of Physics and Astronomy, Carleton College,
Northfield, MN 55057}

\begin{abstract}
We report on preliminary results from the recent multi-epoch neutral 
hydrogen absorption measurements toward three pulsars, B0823+26, B1133+16 and 
B2016+28, using the Arecibo telescope. 
We {\em do not} find significant variations in optical depth
profiles over periods of 0.3 and 9--10 yr, or on spatial 
scales of 10--20 and 70--85 AU. 
The large number of non detections of the tiny scale atomic structure
suggests that the AU-sized structure is not ubiquitous in the 
interstellar medium and could be quite a rare phenomenon.
\end{abstract}


\section{Introduction}

For many years, both observations and theory have  provided extensive 
support for the existence of structure in the interstellar medium (ISM) on 
scales from $\sim$ 1 kpc down to $\sim$ 1 pc  (cf. \cite{Dickey90}).
However, it has long been expected that structure on smaller scales 
($<1$ pc) would {\em{not}} be prominent in the ISM \citep{Heiles00}. 
Indeed, using  median values for thermal pressure and temperature of the cold 
neutral medium (CNM) of  $P_{\rm th} \sim 2250$ cm$^{-3}$ K
\citep{Jenkins01} and T $\sim$70 K \citep{Heiles03b},
the expected volume density for the CNM clouds is about 30
cm$^{-3}$.  The median measured column density of 5$\times10^{19}$
cm$^{-2}$ \citep{Heiles03b}, would then indicate that the {\it typical} 
expected scale length for the CNM features is $\sim1$ pc, in conformance
with the standard theory and observations.   

Consequently, it was quite surprising when observers began to find 
structure on $10^{1-2}$ AU scales in many different directions in the ISM.  
The apparent detections of the AU-sized structure in the cold neutral 
atomic hydrogen (HI) medium, called tiny--scale 
atomic structure [``TSAS'', \cite{Heiles97}], were obtained
with the following techniques:
(1) spatial mapping  of HI absorption line profiles across 
extended background sources \citep{Dieter76,Diamond89,Davis96,Faison01}; 
(2) temporal and spatial variations of optical interstellar 
absorption lines against binary stars and globular clusters 
\citep{Meyer96,Meyer99}; and  
(3) time variability of HI absorption profiles against pulsars
\citep{Deshpande92,Frail94,Johnston03}.  
Since the AU-sized structure was detected very frequently, it was thought
that it is likely to be a general property of the ISM.

TSAS observations  yield measurements of optical depth variation 
($\Delta \tau$) and the particular temporal baseline over 
which the variation is measured. The baseline can also be converted  
into a transverse angular size ($L_{\perp}$) if the background source 
speed or distance is known, respectively. 
A straightforward  interpretation is that the 
optical depth variations are due to a blob, whose transverse
dimension is equal to $L_{\perp}$,  moving into and out of the line-of-sight. 
Assuming a simple spherical geometry, the HI volume density of these blobs 
can be estimated and is typically of order of $10^{4}$ cm$^{-3}$, far 
above the canonically expected value. The inferred thermal pressure is 
then of the order of $10^{6}$ cm$^{-3}$ K (assuming T $\sim$40--70 K), 
which is also much higher than the hydrostatic equilibrium pressure of 
the ISM or the standard thermal pressure of the CNM. 
These problems have been known for a long time and 
have caused much controversy. It is expected that such over-dense and 
over-pressured features should dissipate on a time scale of about 100 yrs 
and should therefore not be common in the ISM, yet the TSAS observations 
have indicated that they are apparently ubiquitous. 
Furthermore, \cite{Jenkins01} found evidence for the existence of
over-pressured gas ($P/k \ga 10^{5}$ cm$^{-3}$ K) using CI fine-structure 
excitation, indicating that pressure enhancements could be widespread in
the ISM and may even be related to the TSAS.

Several explanations were proposed to reconcile the observations of 
extensive  TSAS with theory. \cite{Heiles97} proposed that
TSAS features are curved filaments and/or sheets that happen to
be aligned along our line-of-sight. \cite{Deshpande00a} suggested that 
TSAS blobs correspond to the tail of a hierarchical structure organization that
exists on larger scales, rather than discrete structures with longitudinal
dimension $L_{\perp}$ and  extraordinarily high volume densities. 
\cite{Gwinn01} proposed that optical depth fluctuations
seen in multi-epoch pulsar observations are a scintillation
phenomenon combined with the velocity gradient across the absorbing HI.
Different explanations predict a different level
of optical depth variations at a particular scale size. For example,
\cite{Deshpande00a} expects that optical depth variations would increase
with the size of structure, while \cite{Gwinn01} predicts maximum
variations on the very small spatial scales probed by interstellar 
scintillation. All suggested explanations, however, call for more 
observational data. 

Motivated by the recent theoretical efforts in understanding the nature and
origin of the TSAS we have undertaken new multi-epoch observations of HI
absorption against a set of bright pulsars.  We chose to observe the same
sources as Frail et al. (1994) in order to enhance the number of available
time baselines for comparison. This paper summarizes our first results 
from this project for three pulsars, B0823+26, B1133+16 and B2016+28. 
In Section~\ref{s:obs} we summarize our observing and data processing 
strategies. Results on individual objects are presented in 
Section~\ref{s:results} and discussed in Section~\ref{s:discussion}.

\section{Observations and Data Processing}
\label{s:obs}

\begin{figure*}
\epsscale{1.4}
\plotone{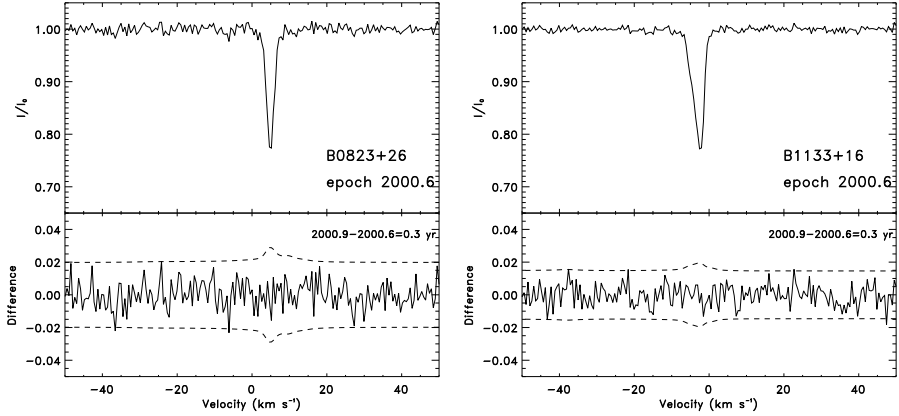}
\caption{\label{f:0823_1133}{\em Top:} HI absorption spectra 
against B0823+26 (left) and B1133+16 (right) observed at epoch
2000.6. {\em Bottom:} Difference between the 
absorption spectra obtained in  2000.6 and 2000.9. 
Contours show $\pm$2-$\sigma$ levels where the HI brightness 
temperature has been taken into account.}
\end{figure*}

We have used the Arecibo telescope\footnote{The Arecibo Observatory
is part of the National Astronomy
and Ionosphere Center, operated by Cornell University under a
cooperative agreement with the National Science Foundation.}
to obtain new multi-epoch HI absorption measurements against six
pulsars previously studied by Frail et. al. (1994).
The observing and data processing procedures were very similar to
\cite{Stanimirovic03}, who measured pulsar OH absorption profiles at 
Arecibo with the same backend, the Caltech Baseband Recorder.
We had four observing sessions: August 2000, December 2000,
September 2001 and November 2001, searching for HI absorption profile 
variations over time intervals from less than a day to 1.25 years. Data from 
Frail et. al. (1994) extend the time baselines to a decade.

Our basic data product was a cube of pulsar intensity as a function of
pulsar rotational phase and radio frequency. `Pulsar-on' and `pulsar-off' 
spectra were then accumulated by finding the pulsar pulse in software, 
and extracting spectra during the pulse and between pulses, respectively. 
The pulsar absorption spectrum is created by generating the `pulsar-on' --
`pulsar-off' spectrum for each scan, doing frequency switching to flatten
the baseline, and accumulating all such spectra with a weight 
proportional to $T^2_{\rm PSR}$, where  $T_{\rm PSR}$ is the antenna
 temperature of the pulsar. Final absorption and emission spectra 
have velocity resolution of 0.5 \kms.

\section{Results on individual sources}
\label{s:results}

\begin{figure*}
\epsscale{1.4}
\plotone{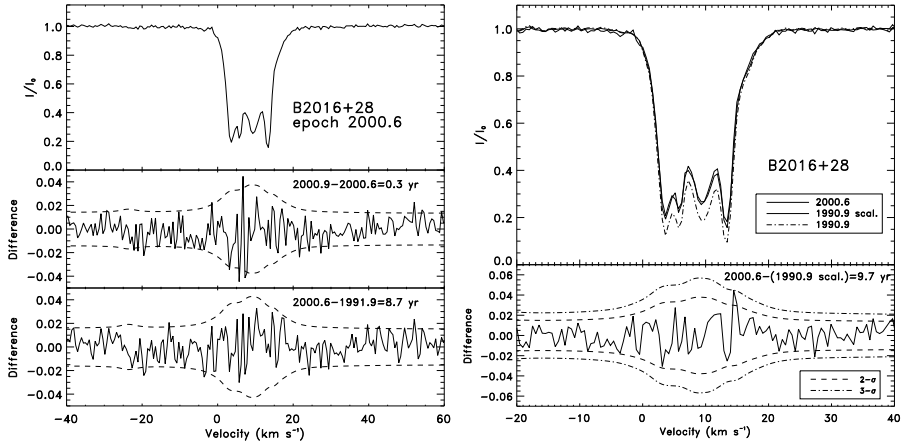}
\caption{\label{f:2016}{\bf Left} ({\it Top}) HI absorption profile 
toward B2016+28 observed at epoch 2000.6. ({\it Middle}) 
Difference between absorption profiles obtained in 2000.6 and 2000.9.
({\it Bottom}) Difference between absorption profiles obtained in 2000.6 and
1991.9 (from Frail et al. 1994). Contours show $\pm$2-$\sigma$ levels.
{\bf Right} ({\it Top}) The HI absorption profile toward B2016+28 obtained
in 2000.6 is shown with a thin solid line. The HI absorption spectrum from 
1990.9 from Frail et al. (1994) is shown with a dot-dashed line; the same
spectrum multiplied by 0.9 and after adding an offset of 0.1 is shown 
as a thick solid line.
({\it Bottom}) Difference between the 2000.6 profile and the scaled 
1990.9 profile, 
with contours showing $\pm$2-$\sigma$ (dashed line) and $\pm$3-$\sigma$ 
(dot-dashed line) significance levels.}
\end{figure*}

For each pulsar, we display a single absorption spectrum on top, and then
one or more absorption  spectrum differences between two epochs.   
Overlaid atop each difference spectrum is a  $\pm$2-$\sigma$ significance 
envelope. To calculate this expected  noise level,  
we have taken into account the following contributions.
(1) The sky background contribution was estimated from the all sky survey 
at 408 MHz by \cite{Haslam82} and scaled to 1.4 GHz using a spectral index 
of $-2.6$. (2) The contribution from HI {\em{emission}} was estimated from
the `pulsar-off' spectra and scaled appropriately to match previously published
observations. The effect of this contribution can be
very significant. For example the rms noise on--line is three times higher
than the rms noise off-line for the case of B2016+28.
(3) The  pulsar continuum emission itself is also  sufficiently strong to  
contribute an additional  $\sim10-30\%$ to the noise temperature. 
Information on individual objects and an upper limit 
on $\Delta\tau$ are given in Table 1 (note that these are 2-$\sigma$ limits 
at the HI line). Our measurements and the  Frail et al. (1994) results are
discussed for each source below.

\subsection{PSR B0823+26}

Fig.~\ref{f:0823_1133} (left) compares HI absorption profiles
obtained toward B0823+26 in 2000.6 and 2000.9.
We find {\em{no}} significant change in absorption  over this time span
down to a $\Delta\tau$ level of about 0.04. The time interval of 0.3 yr
translates to a transverse distance of about 10 AU.
Frail et al. (1994) also found almost no variations over a period of 0.6 yr,
but {\em{did}} report variations of about 0.07 over a period of 1.1 yr. See
the Discussion section below for an examination of the discrepancies 
between the two groups' results.  

\subsection{PSR B1133+16}

Fig.~\ref{f:0823_1133} (right) displays HI absorption profiles
obtained toward B1133+16  in 2000.6 and 2000.9.
We detect {\em{no}} variations down to a $\Delta\tau$ level of about
0.02. During this period, the pulsar traveled 20 AU.  Frail et al. (1994) 
also  detected no significant variations on their 0.6 yr baseline on 
this pulsar, but {\em{did}} see variations in $\tau$ of about 0.04 over
a period of 1.1 yr. We investigate the discrepancy below.

\subsection{PSR B2016+28--Short and long timescales}

\begin{table*}
\caption{Transverse scales and maximum $\Delta \tau$ probed by the three
pulsars. Note that upper limits on $\Delta \tau$ correspond to 
2-$\sigma$ noise on the HI line while the two values of $\Delta \tau$ that
are not upper limits lie beyond the 2-$\sigma$ noise. 
The longest spatial baseline for B2016+28
corresponds to the difference between the 2000.6 spectrum and the 1990.3
spectrum from Frail et al. (1994) and is not shown in the paper.
Transverse velocities used to calculate $L_{\perp}$ are given in Frail et
al. (1994).}
\centering
\label{table1}
\begin{tabular}{lll}
\noalign{\smallskip} \hline \hline \noalign{\smallskip}
PSR      & $L_{\perp}$ (AU) & Max. $\Delta \tau$ \\
\hline
B0823+26 & 10        & $<0.04$ \\
B1133+16 & 20        & $<0.03$    \\
B2016+28 & 3         & ~~~0.18    \\
         & 70        & $<0.20$    \\
         & 80        & ~~~0.15    \\
         & 85        & $<0.20$    \\
\noalign{\smallskip} \hline \noalign{\smallskip}
\end{tabular}
\end{table*}

For this pulsar, we have compared HI absorption profiles
on our usual short ($\sim0.3$ yr) timescale, but also on
decade--long scales. On the left side of Fig.~\ref{f:2016}, in the 
middle panel we 
show the usual difference of absorption spectra from 2000.6 and 2000.9, 
while the bottom left panel investigates an 8.7 year interval by 
differencing our 2000.6 spectrum and the Frail et al. (1994) 1991.9 data.  
The 0.3 yr  baseline, corresponding to a $\sim 3$ AU scale, shows only a 
marginal 2.6-$\sigma$ change in absorption, while the 8.7 yr,
$\sim 70$ AU baseline, does not exhibit any
significant variations at all. This result is very different from
Frail et al. (1994), who found very large optical depth variations of 
$\Delta\tau \ga 1$  over periods of 0.6 and 1.7 yr.

The  Frail et al. (1994) epoch 1991.9 absorption spectrum that we use
for the above comparison is particularly different from the previous 
three Frail et al. epochs (their first one actually being from
\cite{Clifton88}). While it agrees  well with our epoch $\sim2000$ results, 
their 1991.9  profile exhibits significantly shallower 
absorption in all four principal absorption features than do their earlier
epoch results.  Frail et al. (1994) remarked that the apparent change 
could be caused by an incorrect normalization but were confident that 
was not the case.

We further investigate the apparent Frail et al. variations by 
comparing our 2000.6 spectrum with the pre-1991.9 
Frail et al. profiles. As expected, there is a large 
difference in the depth of all four absorption features, as shown in 
the top plot on the right side of Fig.~\ref{f:2016} 
where the 1990.9 Frail et al.  spectrum is plotted with a dot-dashed 
line and our 2000.6 result is plotted as a thin solid line.  
Since all four lines exhibit the same trend, it
seems reasonable  that the differences may result from a slight 
calibration problem in the Frail et al. data. To test this hypothesis,
we performed a least--squares fit for a single scale factor (plus an
offset) that would minimize the difference between these two spectra.

After applying this fitted scale factor ($0.900\pm0.005$) and a constant 
($0.097\pm0.004$) to the 1990.9 spectrum, we arrive at the ``scaled'' 
1990.9 spectrum, shown as a 
thick solid line in the top right plot of Fig.~\ref{f:2016}. Note that
this process brings the previously discrepant 1990.9 profile into 
excellent agreement with the 2000.6 result.  The bottom right plot  
of Fig.~\ref{f:2016} emphasizes this result by exhibiting the difference
between the 2000.6 and scaled 1990.9 spectra. A single 3-$\sigma$ deviation
is all that remains.  Note that without scaling, this change would be 
much larger. We conclude that most of the apparent variations among 
the Frail et al. B2016+28 spectra, and
hence between our epoch 2000.6 data and the early Frail et al. epochs,
could result from a systematic calibration issue in the Frail et al. data.
The measurement of pulsar absorption spectra is a very delicate process
fraught with subtle issues related to the strongly varying pulsar signal.
We speculate that the three-level correlation spectrometer used by Frail 
et al. was more prone to calibration problems than our current four-level
spectrometer.

\section{Discussion}
\label{s:discussion}

Our observations  probe the existence of temporal variations in HI
absorption profiles which would be indicative of the existence of the TSAS on
spatial scales of several tens of AU in several different directions. 
The canonical TSAS has $L_{\perp}=30$ AU, extracted by Heiles (1997) from
previous observations, meaning that TSAS should be common at the scales
that we have studied. Yet in all of the short and long baselines we have
investigated, we find only two {\em marginal} detections of changing absorption.

Our sensitivity is better than any level of variations reported 
previously \citep{Frail94,Johnston03}, at velocity resolution 
of 0.5 \kms, meaning that our non-detections are a significant result.  
In addition, the upper limits on optical depth fluctuations, 
set by B0823+26 and B1133+16 at scales of 10--20 AU,
$\Delta\tau=0.03-0.04$, are strikingly low. 
A $\Delta\tau$ of 0.03 corresponds to $5\times10^{18}$ cm$^{-2}$ for 
the column density fluctuations of the CNM, for assumed spin temperature of
50 K and the full-width half-maxima of absorption features of 2 \kms. 
On slightly larger scales of 70--80 AUs probed by B2016+28 over periods of
almost a decade we find a single 3-$\sigma$ spike of $\Delta\tau=0.15$, 
a $\sim2.5$-$\sigma$ feature of $\Delta\tau \sim 0.18$, and non-detections 
in all other cases.

Deshpande (2000) predicts opacity variations as an extension of HI opacity
irregularities observed on larger scales using a single power law spectrum.
Using the power spectrum of opacity distribution in the direction of Cas A
with the power-law index of 2.75 and extrapolating down to AU scales, 
they predict  $\Delta\tau <0.1$ at scales $\la20$ AU (from their Fig. 2), 
while at 50--100 AU $\Delta\tau \sim$0.2--0.4 is expected.  
This model generally expects that $\Delta\tau$ increases with spatial
scales. Our results for all 
three pulsars seem consistent with this picture so far.

Why are the (primarily) non--detections  presented here and by Johnston 
et al. (2003) so different from the results of Frail et al. (1994)?  
We showed above that in the case of B2016+28 where multiple 
absorption features are visible it is possible to simply scale 
absorption spectra from different epochs and get rid of most of the 
variations, indicating that slight calibration problems may be
responsible. 
A similar scaling technique could significantly reduce variations
seen in some of the other Frail et al.'s pulsars 
(e.g. B0823+16, B1133+6, B1737+13). However, 
it is not obvious that this scaling could be justified since
only single principal absorption features are visible in the spectra.
In addition, Johnston et al. (2003) have already noted that 
many of the differences noted by Frail et al. may not be as significant 
as they stated.

This large number of TSAS non detections, together with recent results by
Johnston et al. (2003), is disturbing and unexpected!  While previous
observations were frequently detecting TSAS, in six comparisons presented
here, and in many comparisons in Johnston et al. (2003), there are only two
{\em marginal} detections for the case of B2016+28 and one detection for B1641-45
(Johnston et al. 2003) over a period of almost twenty years.  This raises
the question of the existence of TSAS as traced at least by multi-epoch
pulsars observations. Contrary to the previous belief that the TSAS is
ubiquitous in the ISM, our results indicate that TSAS is rather rare and
could be related to some kind of isolated, sporadic events
such as superbubble or supernova expansion, stellar mass ejecta, or 
a bow-shock propagation.

\section{Conclusions}

We have compared multi-epoch HI absorption observations toward B0823+26,
B1133+16 and B2016+28 over short and long periods. Except for 
two marginal changes in the case of B2016+28, we {\em do not} find 
significant changes in absorption spectra. This is very different from 
previous observations by Frail et al. (1994) who saw significant 
optical depth variations for the same pulsars over time periods 
of 1.1 and 1.7 yrs. 
We have shown that in the case of B2016+28 most of the variations in
absorption profiles of Frail et al. (1994) could be due to a systematic
calibration problem.
A large number of non detections of the TSAS presented here, together
with recent results by Johnston et al. (2003), suggests that the TSAS is
not ubiquitous in the ISM and could be a rare phenomenon.

\begin{acknowledgements}
We are grateful to Caltech's Center for Advance Computation and Research
for the use of their facilities for data storage and processing.
We express our thanks to Stuart Anderson and Rick Jenet for help 
with observations and data processing. We thank Carl Heiles for
stimulating discussions, and an anonymous referee 
for insightful suggestions.
SS acknowledges support by NSF grants  AST-0097417 and AST-9981308.
KED, AH, JTG, and JMW were supported by NSF Grant AST-0098540.
\end{acknowledgements}


\label{lastpage}
\end{document}